\documentclass[pre,twocolumn]{revtex4-2}
\usepackage{amsmath}
\usepackage{amssymb}
\usepackage{graphicx}
\usepackage{braket}
\usepackage{stmaryrd}
\usepackage{hyperref}
\usepackage{color}
\usepackage{dsfont}
\usepackage{bm}
\usepackage{wasysym}
\usepackage{mathrsfs}
\usepackage{booktabs}

\hypersetup{
    colorlinks,
    citecolor=blue,
    linkcolor=blue,
    urlcolor=blue
}

\begin{document}

\title{
Statistical Decision Theory in Thermodynamics
}
\author{Yoshihiko Hasegawa}
\email{hasegawa@biom.t.u-tokyo.ac.jp}
\affiliation{Department of Electrical Engineering and Information Systems, Graduate School of Engineering, The University of Tokyo,
Tokyo 113-8656, Japan}

\date{\today}
\begin{abstract}

We introduce statistical decision theory in stochastic thermodynamic systems. Statistical decision theory integrates several statistical tasks, such as estimation, hypothesis testing, and classification. 
In this study, we consider statistical decision theory in stochastic thermodynamics, which unifies sensing, classification, and uncertainty relations.
In particular, we focus on the minimax risk, the smallest achievable expected loss under the worst-case parameter choice, even with an optimal estimator.
We derive lower bounds on the minimax risk using Le Cam's two-point method. 
The derived relationships can be used across a broader range of thermodynamic scenarios.
To illustrate these relations, we study classification and distribution estimation problems in a Markov process. We demonstrate that the minimax risk provides lower bounds given by the dynamical activity in these problems.
Furthermore, by examining particular dynamics, we establish that the resulting bounds simplify to uncertainty relations involving dynamical activity or entropy production.

\end{abstract}
\maketitle

\section{Introduction\label{sec:introduction}}

\begin{figure}
\centering
\includegraphics[width=1\linewidth]{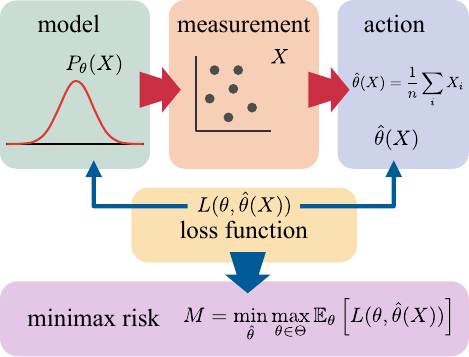}
\caption{
Schematic illustration of statistical decision theory.
There is a true parameter $\theta$, and a distribution $P_\theta(X)$ parameterized by it.
Observations from this distribution yield a measurement $X$.
Based on this measurement, an action $\hat{\theta}(X)$ is selected.
The loss between $\theta$ and $\hat{\theta}(X)$ is represented by $L$, and the loss associated with the action that minimizes the worst-case expected loss is called the minimax risk $M$.
}
\label{fig:SDT}
\end{figure}

Parameter inference plays a central role in statistics and machine learning.
A central object of interest in parameter estimation is how precisely an estimator can infer a parameter. 
One major answer to this was formulated by the Cram\'er-Rao inequality \cite{Cramer:1946:Book,Rao:1945:Bound}, which states that the variance of an unbiased estimator
is bounded from below by the inverse of the Fisher information. 
The importance of the Cram\'er-Rao inequality is growing in nonequilibrium thermodynamics \cite{Seifert:2012:FTReview,VandenBroeck:2015:Review}.
In particular, thermodynamic uncertainty relations \cite{Barato:2015:UncRel,Gingrich:2016:TUP,Garrahan:2017:TUR,Dechant:2018:TUR,Terlizzi:2019:KUR,Hasegawa:2019:CRI,Hasegawa:2019:FTUR,Timpanaro:2019:EFTTUR,Dechant:2020:FRIPNAS,Vo:2020:TURCSLPRE,Koyuk:2020:TUR,Saryal:2019:TUR,Prech:2024:ClockUR}, which are a central topic in nonequilibrium thermodynamics, 
can be regarded as a thermodynamic implementation of the Cram\'er-Rao inequality. 
Although thermodynamic uncertainty relations have been applied to several systems, including quantum dynamics \cite{Hasegawa:2020:QTURPRL,Yunoki:2025:FeedbackQTUR,Yunoki:2026:HeisenbergScaling}, 
they cannot account for more complicated problems.
Ultimately, this is because the scope of problems to which the framework of statistical inference can be applied is limited.
In statistics, there are problems that cannot be addressed within the framework of parameter estimation; however, the framework known as statistical decision theory \cite{LeCam:1973:dimensionality,Raginsky:2013:SLT,Wainwright:2019:HighDimensionalStatistics} is widely used to tackle these problems.
Statistical decision theory, first introduced by Wald \cite{Wald:1949:Decision}, integrates several statistical and machine learning tasks, such as
estimation, hypothesis testing, classification, and so on.
In particular, for problems such as learning, regression, and classification, where errors cannot be simply evaluated by the estimator variance, statistical decision theory has been extensively studied. 
Statistical decision theory is an approach that determines the optimal action given data in environments with uncertainty (Fig.~\ref{fig:SDT}).
Statistical decision theory introduces a loss function and finds an optimal action based on the loss function. 
For example, from the viewpoint of statistical decision theory, parameter estimation is the problem of determining a parameter of interest from data, where the loss function is the mean squared error. 
While the Cram\'er-Rao inequality only quantifies the uncertainty of an unbiased estimator via its variance, statistical decision theory can handle a more general class of loss functions.

In this study, we consider statistical decision theory in stochastic thermodynamics, which we refer to as \textit{thermodynamic statistical decision theory}.  
Thermodynamic statistical decision theory integrates sensing, classification, and uncertainty relations in thermodynamic scenarios.
Specifically, we consider the minimax risk, which is the expected loss that cannot be avoided even with optimal estimation, assuming the worst-case scenario (the most unfavorable parameter). 
We derive lower bounds on the minimax risk via an approach known as Le Cam's two-point method. 
As mentioned above, statistical decision theory provides a unified framework for several statistical problems and can be applied to a wider variety of problems in statistics.
Similarly, thermodynamic statistical decision theory can be applied to a wider range of thermodynamic situations. 
As an application of the bound, we consider classification based on a Markov process. 
We show that the minimax risk, which corresponds to the worst-case misclassification probability, is bounded from below by the sum of dynamical activities. 
Moreover, we show that the derived bound reduces to uncertainty relations comprising dynamical activity or entropy production by considering specific dynamics.
Although several uncertainty relations have hitherto been derived,
one of the advantages of deriving uncertainty relations via thermodynamic statistical decision theory is that 
we can employ general loss functions beyond squared loss.

\section{Methods\label{sec:methods}}

In this study, we consider statistical decision theory in thermodynamics.
First, we review the basics of statistical decision theory \cite{Lehmann:1998:PointEstimation,Kubokawa:2022:foundations}.

Let $\theta$ be a parameter, and let the set of all possible values of $\theta$ be denoted by $\Theta$, which is called the parameter space.
Here, $\theta$ can be a scalar or a vector. 
Let $X$ be the observation.
Let $P_\theta(X)$ be a probability distribution of the observation $X$, which is parametrized by the parameter $\theta$. 
The goal of statistical decision theory is to determine an action from the given observation $X$.
For example, when we are considering parameter estimation from the observation, the action corresponds to an estimator $\hat{\theta}(X)$. 
Although other types of actions can be considered in statistical decision theory, we restrict our attention to an estimator $\hat{\theta}(X)$. 
$\hat{\theta}(X)$ can take values in $\Theta$, i.e., $\hat{\theta}(X)\in \Theta$.

To determine the optimality of the estimator $\hat{\theta}(X)$, we define a loss function $L(\cdot,\cdot)$. 
The loss function $L(\theta,\theta^\prime)$ quantifies the discrepancy between two parameters $\theta$ and $\theta^\prime$. 
We demand that the loss function satisfy the following conditions:
\begin{align}
L(\theta,\theta^{\prime})&=0\,\,\text{if and only if}\,\,\theta=\theta^{\prime},\label{eq:metric_zero_condition}\\L(\theta,\theta^{\prime})&=L(\theta^{\prime},\theta)\ge0,\label{eq:pseudo_metric_condition2}\\\gamma L(\theta,\theta^{\prime})&\le L(\theta,\theta^{\prime\prime})+L(\theta^{\prime\prime},\theta^{\prime}),\label{eq:pseudo_metric_condition1}
\end{align}
where $\gamma > 0$. 
A metric satisfying Eqs.~\eqref{eq:metric_zero_condition}--\eqref{eq:pseudo_metric_condition1} is known as a $b$-metric
\cite{Czerwik:1993:b-metric,Roshan:2015:SuzukiFixedPoint}. 
When $\gamma = 1$, 
the $b$-metric reduces to the conventional metric;
the condition of Eq.~\eqref{eq:pseudo_metric_condition1} represents the triangle inequality. Therefore, $L(\theta,\theta^{\prime})=|\theta-\theta^{\prime}|$ satisfies Eq.~\eqref{eq:pseudo_metric_condition1} with $\gamma = 1$.
The most important instance of the $b$-metric is $|\theta-\theta^\prime|^p$, where $p>0$. 
From Ref.~\cite{Kirk:2013:metric-transforms}, the following relation holds:
\begin{align}
    \left|\theta-\theta^{\prime}\right|^{p}\le\left|\theta-\theta^{\prime\prime}\right|^{p}+\left|\theta^{\prime}-\theta^{\prime\prime}\right|^{p}\,\,\,(0<p<1).
    \label{eq:snowflake}
\end{align}
For $p>1$, it is known that \cite{Roshan:2015:SuzukiFixedPoint}
\begin{align}
    2^{1-p}\left|\theta-\theta^{\prime}\right|^{p}\le\left|\theta-\theta^{\prime\prime}\right|^{p}+\left|\theta^{\prime}-\theta^{\prime\prime}\right|^{p}\,\,\,(p>1).
    \label{eq:bmetric_triangle}
\end{align}
Therefore, the squared loss $L(\theta,\theta^{\prime})=(\theta-\theta^{\prime})^{2}$
is a $b$-metric with $\gamma = 1/2$. 
Strictly speaking, the first condition in Eq.~\eqref{eq:metric_zero_condition} is not necessary when deriving Le Cam's two-point method as shown below. However, this condition will be used when considering consequences of Le Cam's bound in thermodynamic scenarios.

Since the observation $X$ is a random variable, we evaluate the expectation value of the loss function
\begin{align}
    R(\theta,\hat{\theta})\equiv\mathbb{E}_{\theta}[L(\theta,\hat{\theta}(X))]=\sum_{X}P_\theta(X)L(\theta,\hat{\theta}(X)).
    \label{eq:risk_function_def}
\end{align}
The function $R$ is called the risk function. We aim to determine the estimator $\hat{\theta}$ via the risk function $R$. 
The estimator attempts to estimate $\theta$ from the observation $X$.

The estimator $\hat{\theta}(X)$ is determined so that the risk function $R$ becomes minimal. 
When minimizing the risk function with respect to the estimator $\hat{\theta}(X)$, we need to specify $\theta$ in Eq.~\eqref{eq:risk_function_def}. 
Importantly, the risk function is a function of $\theta$. Suppose there are two estimators $\hat{\theta}_1$ and $\hat{\theta}_2$. Even if the relation $R(\theta,\hat{\theta}_1)<R(\theta,\hat{\theta}_2)$ holds
for a parameter $\theta$, the inequality may be $R(\theta^\prime,\hat{\theta}_1)>R(\theta^\prime,\hat{\theta}_2)$ for another parameter $\theta^\prime$. This simple example shows that the best estimator depends on a chosen $\theta$ and cannot be simply determined. 
A common choice is to consider the worst-case scenario.
In this case, the risk function is given by
\begin{align}
    M=\min_{\hat{\theta}}\max_{\theta\in\Theta}R(\theta,\hat{\theta}).
    \label{eq:minimax_risk}
\end{align}
$M$ is referred to as the \textit{minimax risk}. 
The estimator determined via Eq.~\eqref{eq:minimax_risk} is known as the minimax estimator. 
Given an estimator $\hat{\theta}(X)$, the worst risk is $\max_{\theta \in \Theta} R(\theta,\hat{\theta})$.
The minimax risk $M$ is the minimum of such worst-case risks. 

Because the evaluation of the minimax risk $M$ is difficult, several approximation schemes are used. 
The most common approach is Le Cam's two-point method. 
Le Cam's two-point method gives a lower bound on the minimax error \cite{LeCam:1973:dimensionality} (also see \cite{Tibshirani:2024:statlearn,Kim:2020:minimaxlowerbounds,Wainwright:2019:HighDimensionalStatistics,Raginsky:2013:SLT}). This method is based on the following simple observation:
\begin{align}
    \max_{\theta\in\Theta}R(\theta,\hat{\theta})\ge\max_{\theta\in\{\theta_{0},\theta_{1}\}}R(\theta,\hat{\theta}),
    \label{eq:two_theta_lowerbound}
\end{align}
where $\theta_0,\theta_1 \in \Theta$.
Equation~\eqref{eq:two_theta_lowerbound} yields
\begin{align}
    M\geq\min_{\hat{\theta}}\max_{\theta\in\{\theta_{0},\theta_{1}\}}R\left(\theta,\hat{\theta}\right).
    \label{eq:M_lowerbound_Le_cam}
\end{align}
Equation~\eqref{eq:M_lowerbound_Le_cam} reduces the original maximization over the parameter set $\Theta$ to only two parameters $\theta_0$ and $\theta_1$ in the set. 
Then Le Cam's two-point method provides the following lower bound \cite{LeCam:1973:dimensionality,Wainwright:2019:HighDimensionalStatistics} (see Appendix~\ref{sec:lecam_derivation} for a derivation):
\begin{align}
    M\geq\frac{1}{2}\gamma L\left(\theta_{0},\theta_{1}\right)\left(1-\mathrm{TV}(P_{\theta_{0}},P_{\theta_{1}})\right).
    \label{eq:LeCam_two_point}
\end{align}
Here, $\mathrm{TV}(\cdot,\cdot)$ is the total variation distance between two probability distributions $P(x)$ and $Q(x)$:
\begin{align}
    \mathrm{TV}(P,Q)\equiv\frac{1}{2}\sum_{x}|P(x)-Q(x)|.
    \label{eq:TV_distance_def}
\end{align}
Equation~\eqref{eq:LeCam_two_point} is extensively used in the field of statistical decision theory \cite{Huang:2025:LeCam}. 
Equation~\eqref{eq:LeCam_two_point} provides the lower bound in terms of the total variation distance. 
It is often the case that the lower bound comprises another information quantity. The Bhattacharyya coefficient is defined by
\begin{align}
    \mathrm{Bhat}(P,Q)=\sum_{x}\sqrt{P(x)Q(x)}.
    \label{eq:Bhat_coef_def}
\end{align}
It is known that the following relation holds:
\begin{align}
    \mathrm{TV}(P,Q)\leq\sqrt{1-\mathrm{Bhat}(P,Q)^{2}}.
    \label{eq:TV_and Bhat_ineq}
\end{align}
Using Eq.~\eqref{eq:TV_and Bhat_ineq} in Eq.~\eqref{eq:LeCam_two_point}, the following relation holds:
\begin{align}
    M\geq\frac{1}{2}\gamma L\left(\theta_{0},\theta_{1}\right)\left(1-\sqrt{1-\mathrm{Bhat}(P_{\theta_{0}},P_{\theta_{1}})^{2}}\right).
    \label{eq:M_lowerbound_Bhat}
\end{align}
Instead of using the Bhattacharyya coefficient, the lower bound can be expressed in terms of the Kullback-Leibler divergence via the Pinsker inequality \cite{Pinsker:1964:Book}. 
The use of other divergences will be discussed later.

\section{Results\label{sec:results}}

In this study, we model statistical decision through a continuous-time Markov process. 
Let $\mathfrak{B}=\{1,2, \cdots, D\}$ be a state space of the Markov process. 
Let $W_{\nu\mu}$ be the transition rate of the process from $\mu$ to $\nu$. 
The probability distribution at time $t$ is denoted by $P(\nu;t)$. The time evolution of the Markov process is governed by
\begin{align}
    \frac{d}{d t} P(\nu ; t)=\sum_\mu W_{\nu \mu} P(\mu ; t),
    \label{eq:master_equation}
\end{align}
where the diagonal elements are $W_{\mu \mu}=$ $-\sum_{\nu \neq \mu} W_{\nu \mu}$.
We introduce thermodynamic quantities of the Markov process. 
In stochastic thermodynamics, trajectories play a fundamental role in describing physical quantities at the trajectory level. Let us focus on the dynamics within the time interval $[0,\tau]$, where $\tau>0$. Let $J$ be the number of jumps within $[0,\tau]$. Let $t_j$ be the time of the $j$th jump and let $Z_j$ be the state after the $j$th jump.  
The trajectory is described by
\begin{align}
    \Gamma_\tau=\left[\left(t_0, Z_0\right),\left(t_1, Z_1\right), \ldots,\left(t_J, Z_J\right)\right],
    \label{eq:trajectory_def}
\end{align}
where $t_0=0$ and $Z_0$ is the initial state of the process.  
We can think of the probability distribution of $\Gamma_\tau$:
\begin{align}
    \mathcal{P}[\Gamma_{\tau}]=P(Z_{0};0)\mathcal{P}[\Gamma_{\tau}|Z_{0}].
    \label{eq:path_integral}
\end{align}
In stochastic thermodynamics, several physical quantities come into play. 
The time-integrated entropy production during $[0,\tau]$ is defined by
\begin{align}
    \Sigma(\tau) & \equiv \int_0^\tau \sum_{\mu \neq \nu} W_{\mu \nu} P(\nu ; t) \ln \frac{W_{\mu \nu} P(\nu ; t)}{W_{\nu \mu} P(\mu ; t)} d t\nonumber\\
& =\Delta S+\Delta S_{\mathrm{m}}.
\label{eq:entropy_production_def}
\end{align}
Here, $\Delta S$ is the Shannon entropy difference and $\Delta S_\mathrm{m}$ is the medium entropy increase during $[0,\tau]$. 
Next, we introduce the dynamical activity, which is another widely used quantity in stochastic thermodynamics.
The time-integrated dynamical activity during $[0,\tau]$ is defined by
\begin{align}
    \mathcal{A}(\tau) \equiv \int_0^\tau \mathfrak{a}(t) dt ,
    \label{eq:mathcal_A_def}
\end{align}
where $\mathfrak{a}(t)$ is the dynamical activity:
\begin{align}
    \mathfrak{a}(t) \equiv \sum_{\mu \neq \nu} W_{\mu \nu} P(\nu ; t).
    \label{eq:mathfrak_a_def}
\end{align}
$\mathfrak{a}(t)$ quantifies the average jump frequency at time $t$ and thus quantifies the activity of the system of interest. 
Dynamical activity plays a central role in nonequilibrium thermodynamics \cite{Garrahan:2017:TUR,Shiraishi:2018:SpeedLimit,Terlizzi:2019:KUR,Hasegawa:2023:BulkBoundaryBoundNC}. 

Let us consider a statistical decision for the Markov process. Suppose that the Markov process encodes a parameter $\theta$ and we wish to construct an estimator $\hat{\theta}(\Gamma_\tau)$, which infers the parameter from a stochastic trajectory $\Gamma_\tau$. 
Again, let $\Theta$ be a set of parameters, $\theta \in \Theta$. 
We define the estimator $\hat{\theta}(\Gamma_\tau)$, which is a function of the trajectory $\Gamma_\tau$. 
Let $\mathcal{P}[\Gamma_\tau]$ be the path probability of a trajectory $\Gamma_\tau$, and we assume that the path probability is parametrized by $\theta$ as $\mathcal{P}_{\theta}[\Gamma_{\tau}]$. 
We employ a loss function $L(\theta,\theta^\prime)$, which is the $b$-metric satisfying the conditions shown in Eqs.~\eqref{eq:metric_zero_condition}-\eqref{eq:pseudo_metric_condition1}. 
The minimax risk is given by
\begin{align}
    M=\min_{\hat{\theta}}\max_{\theta \in \Theta}\mathbb{E}_{\theta}[L(\theta,\hat{\theta}(\Gamma_{\tau}))].
    \label{eq:thermodynamic_minimax_risk}
\end{align}
Using Eq.~\eqref{eq:M_lowerbound_Bhat}, 
Le Cam's bound in the thermodynamic scenario is given by
\begin{align}
    M\geq\frac{1}{2}\gamma L\left(\theta_{0},\theta_{1}\right)\left(1-\mathrm{TV}(\mathcal{P}_{\theta_{0}},\mathcal{P}_{\theta_{1}})\right),
    \label{eq:M_lowerbound_TV_thermo}
\end{align}
where $\theta_0,\theta_1 \in \Theta$. 
Again, from Eq.~\eqref{eq:M_lowerbound_Bhat}, we obtain
\begin{align}
    M\geq\frac{1}{2}\gamma L\left(\theta_{0},\theta_{1}\right)\left(1-\sqrt{1-\mathrm{Bhat}(\mathcal{P}_{\theta_{0}},\mathcal{P}_{\theta_{1}})^{2}}\right),
    \label{eq:thermo_M_lowerbound_Bhat}
\end{align}
where $\theta_0,\theta_1 \in \Theta$. 
Moreover, the lower bound can be expressed in terms of the Kullback-Leibler divergence:
\begin{align}
    M\geq\frac{1}{2}\gamma L\left(\theta_{0},\theta_{1}\right)\left(1-\sqrt{\frac{1}{2}\mathrm{KL}\left(\mathcal{P}_{\theta_{0}}\|\mathcal{P}_{\theta_{1}}\right)}\right).
    \label{eq:thermo_M_KL_lowerbound}
\end{align}
Because the lower bound in Eq.~\eqref{eq:thermo_M_lowerbound_Bhat} is not expressed in terms of a thermodynamic quantity, we employ the following inequality from Ref.~\cite{Hasegawa:2026:classifiers}:
\begin{align}
    \mathrm{Bhat}(\mathcal{P}_{\theta_{0}},\mathcal{P}_{\theta_{1}})^{2}\ge\exp\left[-\tau(\mathfrak{a}_{\theta_{0}}(0)+\mathfrak{a}_{\theta_{1}}(0))\right].
    \label{eq:Bhat_inequality}
\end{align}
Here, $\mathfrak{a}_{\theta_0}(0)$ and $\mathfrak{a}_{\theta_1}(0)$ are dynamical activities [Eq.~\eqref{eq:mathfrak_a_def}] corresponding to $\theta_0$ and $\theta_1$, respectively, at time $t=0$. 
Equation~\eqref{eq:Bhat_inequality} holds when the initial probability distributions for the two processes are the same. 
Using Eq.~\eqref{eq:Bhat_inequality} in Eq.~\eqref{eq:thermo_M_lowerbound_Bhat}, we obtain
\begin{align}
    M\geq\frac{1}{2}\gamma L\left(\theta_{0},\theta_{1}\right)\left(1-\sqrt{1-\exp\left[-\tau(\mathfrak{a}_{\theta_{0}}(0)+\mathfrak{a}_{\theta_{1}}(0))\right]}\right).
    \label{eq:main_result_1}
\end{align}
Equation~\eqref{eq:main_result_1} shows that the minimax risk $M$ is bounded from below by dynamical activities corresponding to $\theta_0$ and $\theta_1$. 
In Eq. \eqref{eq:thermo_M_KL_lowerbound}, the two path probabilities correspond to a physical quantity only in specific cases.
As will be shown later, when the two path probabilities correspond to forward and backward dynamics, the divergence becomes entropy production. On the other hand, Eq.~\eqref{eq:main_result_1} holds for arbitrary $\theta_0$ and $\theta_1$. That is, regardless of what dynamics correspond to $\theta_0$ and $\theta_1$, the lower bound of Eq.~\eqref{eq:main_result_1} is expressed in terms of the physical quantity dynamical activity.

In Eq.~\eqref{eq:thermodynamic_minimax_risk}, the risk is expressed in terms of $\theta$. It is possible to formalize the risk using a trajectory observable. 
Let $\hat{\phi}(\Gamma_\tau)$ be an observable of $\Gamma_\tau$. 
Let $\mathbb{E}_{\theta}[\hat{\phi}]$ be the expectation of $\hat{\phi}$ with respect to $\mathcal{P}_\theta$:
\begin{align}
\mathbb{E}_{\theta}[\hat{\phi}]\equiv\sum_{\Gamma_{\tau}}\mathcal{P}_{\theta}[\Gamma_{\tau}]\hat{\phi}(\Gamma_{\tau}).
    \label{eq:mathcalP_def}
\end{align}
Let $\hat{\phi}(\Gamma_\tau)$ be an estimator of $\mathbb{E}_{\theta}[\hat{\phi}]$, which is a function of $\Gamma_\tau$. 
Instead of estimating the parameter $\theta$ directly, we consider a scenario in which the expectation of $\hat{\phi}$ is to be estimated. 
Then, the worst-case risk is given by
\begin{align}
V=\max_{\theta\in\Theta}\mathbb{E}_{\theta}[L(\mathbb{E}_{\theta}[\hat{\phi}],\hat{\phi}(\Gamma_{\tau}))].
\label{eq:Mprime_def}
\end{align}
In Eq.~\eqref{eq:Mprime_def}, the loss function $L$ is a $b$-metric, as was the case above. 
Following the same procedure as in Eq.~\eqref{eq:main_result_1}, we obtain
\begin{align}
    V\geq\frac{1}{2}\gamma L\left(\mathbb{E}_{\theta_{0}}[\hat{\phi}],\mathbb{E}_{\theta_{1}}[\hat{\phi}]\right)\left(1-\mathrm{TV}(\mathcal{P}_{\theta_{0}},\mathcal{P}_{\theta_{1}})\right),
    \label{eq:M_lowerbound_TV_thermo2}
\end{align}
where $\theta_0,\theta_1 \in \Theta$. 
The derivation of Eq.~\eqref{eq:M_lowerbound_TV_thermo2} is shown in Section~\ref{sec:derivation_M_lowerbound_TV_thermo2}. 
The right-hand side of Eq.~\eqref{eq:M_lowerbound_TV_thermo} does not depend on the estimator $\hat{\theta}$.
Therefore, it is possible to consider $\min_{\hat{\theta}}$ on the left-hand side of Eq.~\eqref{eq:M_lowerbound_TV_thermo}.
In contrast, both sides of Eq.~\eqref{eq:M_lowerbound_TV_thermo2} depend on the observable $\hat{\phi}$, and thus we cannot simply minimize only the left-hand side of Eq.~\eqref{eq:M_lowerbound_TV_thermo2} with respect to the observable $\hat{\phi}$.
The bounds based on other divergences, such as the Bhattacharyya coefficient and Kullback-Leibler divergence, can be derived in the same way as in Eqs.~\eqref{eq:thermo_M_lowerbound_Bhat} and \eqref{eq:thermo_M_KL_lowerbound}. 
Again, using Eq.~\eqref{eq:Bhat_inequality} in Eq.~\eqref{eq:M_lowerbound_TV_thermo2}, we obtain
\begin{align}
    V&\geq\frac{1}{2}\gamma  L\left(\mathbb{E}_{\theta_{0}}[\hat{\phi}],\mathbb{E}_{\theta_{1}}[\hat{\phi}]\right)\nonumber\\&\times\left(1-\sqrt{1-\exp\left[-\tau(\mathfrak{a}_{\theta_{0}}(0)+\mathfrak{a}_{\theta_{1}}(0))\right]}\right).
    \label{eq:main_result_2}
\end{align}

\section{Examples\label{sec:examples}}

\subsection{Classification\label{sec:classification}}

Let us discuss examples of the bounds derived in the previous section. 
Here, we consider a $K$-class classification problem.
A stochastic process is selected from the $K$ classes, and according to it,
a trajectory is generated.
From the generated trajectory, we estimate which class it was generated from.
The parameter $\theta$ indicates which class it is, and
we assume it takes one of the values in $\Theta = \{1,2,\cdots,K\}$.
Let us consider the 0-1 loss function:
\begin{align}
    L(\theta,\hat{\theta}(\Gamma_\tau))=\begin{cases}
1 & \theta\ne\hat{\theta}(\Gamma_\tau)\\
0 & \theta=\hat{\theta}(\Gamma_\tau)
\end{cases}.
\label{eq:zeroone_loss_function}
\end{align}
Apparently, Eq.~\eqref{eq:zeroone_loss_function} satisfies the $b$-metric condition with $\gamma=1$. 
The risk function corresponding to 0-1 loss is the misclassification probability, and the minimax risk corresponds to the worst-case misclassification probability. 
From Eq.~\eqref{eq:main_result_1}, we obtain
\begin{align}
    M\geq\frac{1}{2}\left(1-\sqrt{1-\exp\left[-\tau(\mathfrak{a}_{\theta_{1}}(0)+\mathfrak{a}_{\theta_{2}}(0))\right]}\right).
    \label{eq:classification_minimax_bound}
\end{align}
Here, $\theta_1,\theta_2 \in \Theta$ and $\theta_1 \ne \theta_2$. To obtain the best bound, the right-hand side of Eq.~\eqref{eq:classification_minimax_bound} should be maximized with respect to $\theta_1$ and $\theta_2$. 
Equation~\eqref{eq:classification_minimax_bound} shows that the worst-case misclassification probability, expressed by $M$, is bounded from below by the sum of dynamical activities corresponding to two classes. 
The lower bound on the Bayes risk was recently derived by the author \cite{Hasegawa:2026:classifiers}. 
The central difference from Ref.~\cite{Hasegawa:2026:classifiers} is that, while Ref.~\cite{Hasegawa:2026:classifiers} calculated the lower bound on the Bayes risk only for binary classification, Eq.~\eqref{eq:classification_minimax_bound} provides the minimax risk that holds for an arbitrary number of classes. 

\subsection{Probability Distribution Estimation}
Let us consider the problem of estimating the steady-state probability distribution $P_\mathrm{ss}(\nu)$ from the observation of a trajectory $\Gamma_\tau$ within $[0,\tau]$. 
Estimation of a probability distribution in statistics is a classical problem and is widely studied in the literature
\cite{Kamath:2015:DistributionLearning}. 
Let $\hat{P}_\mathrm{ss}(\Gamma_\tau)$ be an estimator of the steady-state probability distribution given a trajectory $\Gamma_\tau$. 
Let $\mathscr{P}_\mathrm{ss}$ be a set of path distributions, where the dynamics is in a steady state, and let $P_\mathrm{ss}(\mathcal{P})$ be the steady-state distribution of $\mathcal{P} \in \mathscr{P}_\mathrm{ss}$. 
Following the construction in Ref.~\cite{Polyanskiy:2025:InformationTheory}, we consider the $L_2$ loss:
\begin{align}
    L(\hat{P}_{\mathrm{ss}},P_{\mathrm{ss}}(\mathcal{P}))=\|P_{\mathrm{ss}}(\mathcal{P})-\hat{P}_{\mathrm{ss}}(\Gamma_\tau)\|_{2},
    \label{eq:L2_loss}
\end{align}
where $ \| a - b\|_p $ is the $L_p$ loss for some $1 \leq p \leq \infty$
with 
 $\|a\|_p =\left(\sum\left|a_i\right|^p\right)^{\frac{1}{p}}$.
Equation~\eqref{eq:L2_loss} calculates the $L_2$ loss of the estimated steady-state distribution $\hat{P}_\mathrm{ss}(\Gamma_\tau)$ with respect to the true steady-state distribution $P_\mathrm{ss}$. 
The minimax risk is given by
\begin{align}
M=\min_{\hat{P}_{\mathrm{ss}}}\max_{\mathcal{P}\in\mathscr{P}_{\mathrm{ss}}}\mathbb{E}_{\mathcal{P}}[L(\hat{P}_{\mathrm{ss}}(\Gamma_\tau),P_{\mathrm{ss}}(\mathcal{P}))].
\end{align}
Le Cam's two-point method provides
\begin{align}
    M\geq\frac{1}{2}L\left(P_{\mathrm{ss}}\left(\mathcal{P}_{\theta_{0}}\right),P_{\mathrm{ss}}\left(\mathcal{P}_{\theta_{1}}\right)\right)\left(1-\sqrt{\frac{1}{2}\mathrm{KL}\left(\mathcal{P}_{\theta_{0}}\|\mathcal{P}_{\theta_{1}}\right)}\right),
    \label{eq:LeCam_distribution}
\end{align}
where $\mathcal{P}_{\theta_0}$ and $\mathcal{P}_{\theta_1} \in \mathscr{P}_\mathrm{ss}$. 
For simplicity, let us restrict our attention to the two-dimensional case, $D=2$. 
We need to choose $\mathcal{P}_{\theta_0}$ and $\mathcal{P}_{\theta_1}$ to construct a lower bound on the minimax risk. 
$\mathcal{P}_{\theta_0}$ and $\mathcal{P}_{\theta_1}$ are steady-state dynamics whose transition rates are $W^{(\theta_0)}$ and $W^{(\theta_1)}$, respectively.
We choose $\theta_0 = 0$ and $\theta_1 = \theta$ in the parametrized transition rate:
\begin{align}
W(\theta)=\left[\begin{array}{cc}
-1 & 1+\theta\\
1 & -1-\theta
\end{array}\right].\label{eq:W1_perturbed}
\end{align}
$\mathcal{P}_{\theta_0}$ is a 2-state Markov process where all transition rates are unity. The perturbed dynamics $\mathcal{P}_{\theta_1}$ is identical to $\mathcal{P}_{\theta_0}$ except that the single transition rate corresponding to $2\to 1$ is perturbed to $1 + \theta$. 
The steady-state distribution of $\mathcal{P}_{\theta_0}$ is uniform, while the steady-state distribution of $\mathcal{P}_{\theta_1}$ is slightly different from the uniform distribution due to the perturbation. 
The corresponding steady-state distributions are given by
\begin{align}
P_{\mathrm{ss}}(\mathcal{P}_{\theta_{1}})=\left[\frac{1+\theta}{2+\theta},\frac{1}{2+\theta}\right].
\label{eq:perturbed_Pss}
\end{align}
The expected $L_2$ loss for the perturbed steady-state is given by
\begin{align}
L\left(P_{\mathrm{ss}}(\mathcal{P}_{\theta_0}),P_{\mathrm{ss}}(\mathcal{P}_{\theta_1})\right)=\frac{\theta}{\sqrt{2}(\theta+2)}.
\label{eq:expected_L2_loss}
\end{align}
Substituting Eq.~\eqref{eq:expected_L2_loss} into Eq.~\eqref{eq:thermo_M_KL_lowerbound}, we obtain
\begin{align}
    \min_{\hat{P}_{\mathrm{ss}}}\max_{\mathcal{P}\in\mathscr{P}_{\mathrm{ss}}}\mathbb{E}_{\mathcal{P}}\left[L(\hat{P}_{\mathrm{ss}},P_{\mathrm{ss}}(\mathcal{P}))\right]\ge\frac{1}{8\sqrt{\tau}}\,\,\,(\tau\to\infty).
    \label{eq:distribution_bound}
\end{align}
See Appendix~\ref{sec:distribution_derivation_appendix} for details of the derivation. 
From Eq.~\eqref{eq:distribution_bound}, we see that when estimating the steady-state distribution from a trajectory $\Gamma_\tau$, the lower bound of the $L_2$ loss scales as $\tau^{-1/2}$. 
Note that the lower bound in Eq.~\eqref{eq:distribution_bound} originates from the dynamical activity of a $2$-dimensional Markov chain with transition rate $1$.

\subsection{Uncertainty relations\label{sec:uncertainty_relations}}

We consider specific physical examples of the bound expressed in Eq.~\eqref{eq:M_lowerbound_TV_thermo2}.
Equation~\eqref{eq:M_lowerbound_TV_thermo2} leads to thermodynamic bounds on observable statistics.
Such bounds have been one of the central topics in nonequilibrium thermodynamics. 
Let us consider specific cases of the lower bounds $V$. Consider a counting observable $\hat{\phi}_\sharp(\Gamma_\tau)$.
\begin{align}
    \hat{\phi}_{\sharp}(\Gamma_{\tau})=\sum_{\mu,\nu\,(\mu\ne\nu)}C_{\nu\mu}N_{\nu\mu}.
    \label{eq:counting_obs_def}
\end{align}
Here, $N_{\nu\mu}$ is the number of jumps from $\mu$ to $\nu$ within the interval $[0,\tau]$, and $C_{\nu\mu} \in \mathbb{R}$ is its corresponding weight. 
The counting observable is widely employed in stochastic thermodynamics. 
We assume that the parameter $\theta$ is a positive real value, and it enters the transition rate matrix as 
\begin{align}
    W_{\nu\mu}(\theta)= \theta W_{\nu\mu}.
    \label{eq:W_parametrized_theta}
\end{align}
Moreover, let us assume a steady-state condition in the long-time limit $\tau \to \infty$. Then the counting observable satisfies
\begin{align}
    \mathbb{E}_{\theta+\Delta\theta}[\hat{\phi}_{\sharp}]=\left(1+\frac{\Delta\theta}{\theta}\right)\mathbb{E}_{\theta}[\hat{\phi}_{\sharp}],
    \label{eq:phi_sharp_scaling}
\end{align}
and $\mathbb{E}_{\theta=1}[\hat{\phi}_\sharp] = \mathbb{E}[\hat{\phi}_\sharp]$. 
Suppose that the loss function is 
\begin{align}
    L(\theta,\theta^{\prime})=|\theta-\theta^{\prime}|^{p},
    \label{eq:Lp_def}
\end{align}
($p \ge 1$). In this case, it is known that it is a $b$-metric with $\gamma = 2^{1-p}$ (Eq.~\eqref{eq:bmetric_triangle}). 
Then, from Eq.~\eqref{eq:M_lowerbound_TV_thermo2}, we obtain
\begin{align}
    \frac{\mathbb{E}[\left|\mathbb{E}[\hat{\phi}_{\sharp}]-\hat{\phi}_{\sharp}(\Gamma_{\tau})\right|^{p}]}{\mathbb{E}[\hat{\phi}_{\sharp}]^{p}}\ge\frac{p^{p}}{(p+1)^{p+1}\left(\sqrt{\mathfrak{a}_{\mathrm{ss}}\tau}\right)^{p}},
    \label{eq:generalized_KUR}
\end{align}
which holds for $p>1$ and $\tau \to \infty$. 
The derivation of Eq.~\eqref{eq:generalized_KUR} is shown in Appendix~\ref{sec:derivation_gKUR}. 
For $p=2$, Eq.~\eqref{eq:generalized_KUR} reduces to
\begin{align}
    \frac{\mathrm{Var}[\hat{\phi}_{\sharp}]}{\mathbb{E}[\hat{\phi}_{\sharp}]^{2}}\ge\frac{4}{27\mathfrak{a}_{\mathrm{ss}}\tau}.
    \label{eq:generalized_KUR2}
\end{align}
The left-hand side of Eq.~\eqref{eq:generalized_KUR2} is the relative variance. In Refs.~\cite{Garrahan:2017:TUR,Terlizzi:2019:KUR}, the lower bound on the relative variance is given by $1/\mathfrak{a}_\mathrm{ss}\tau$, which indicates that Eq.~\eqref{eq:generalized_KUR2} is weaker. 
The more general the bounds are, the weaker they are.
Because Le Cam's two-point method applies to very general loss functions, its lower bound is known to be more pessimistic.  
In general, Le Cam's two-point method provides a looser bound, as shown in Appendix~\ref{sec:normal_mean}, in which Le Cam's lower bound and the Cram\'er-Rao lower bound are compared for Gaussian mean inference. 

Next, we consider the observable which vanishes for a no-jump trajectory. Let $\Gamma_\varnothing$ be a trajectory which has no jump. The no-jump observable $\hat{\phi}_\circ(\Gamma_\tau)$ satisfies
\begin{align}
    \hat{\phi}_{\circ}(\Gamma_{\varnothing}) = 0.
    \label{eq:nojump_trajectory}
\end{align}
This observable is widely employed in Refs.~\cite{Hasegawa:2020:TUROQS,Hasegawa:2023:BulkBoundaryBoundNC,Hasegawa:2024:ConcentrationIneqPRL}. 
Equation~\eqref{eq:nojump_trajectory} generalizes the counting observable defined in Eq.~\eqref{eq:counting_obs_def} because the counting observable is $0$ when there is no jump. 
From Eq.~\eqref{eq:main_result_2}, we obtain
\begin{align}
    \mathbb{E}[L(\mathbb{E}[\hat{\phi}_{\circ}],\hat{\phi}_{\circ}(\Gamma_{\tau}))]\geq\frac{1}{2}\gamma L\left(0,\mathbb{E}[\hat{\phi}_{\circ}]\right)\left(1-\sqrt{1-e^{-\tau\mathfrak{a}(0)}}\right),
    \label{eq:generalized_point_KUR}
\end{align}
whose derivation is shown in Appendix~\ref{sec:derivation_pointKUR}. 
When the loss function is given by Eq.~\eqref{eq:Lp_def}, we obtain
\begin{align}
    \frac{\mathbb{E}\left[\left|\hat{\phi}_{\circ}(\Gamma_{\tau})-\mathbb{E}[\hat{\phi}_{\circ}]\right|^{p}\right]}{\left|\mathbb{E}[\hat{\phi}_{\circ}]\right|^{p}}\geq\frac{1-\sqrt{1-e^{-\tau\mathfrak{a}(0)}}}{2^{p}}.
    \label{eq:p_norm_lowerbound}
\end{align}
Equation~\eqref{eq:generalized_point_KUR} is an uncertainty relation similar to 
\begin{align}
    \frac{\mathrm{Var}[\hat{\phi}_{\circ}]}{\mathbb{E}[\hat{\phi}_{\circ}]^{2}}\geq\frac{1}{e^{\mathfrak{a}(0)\tau}-1}.
    \label{eq:point_KUR}
\end{align}
It is important to note that Eq.~\eqref{eq:generalized_point_KUR} holds for an arbitrary $b$-norm. At the expense of the generalization, Eq.~\eqref{eq:generalized_point_KUR} is usually weaker than Eq.~\eqref{eq:point_KUR}. 
So far, we have considered bounds comprising the dynamical activity. 
Since the lower bound in Eq.~\eqref{eq:thermo_M_KL_lowerbound} is given by the entropy production, we can obtain bounds involving entropy production. 

Moreover, using the Bretagnolle-Huber inequality \cite{Bretagnolle:1978:estimation} $\mathrm{TV}(P,Q)\leq\sqrt{1-\exp\left(-\mathrm{KL}(P\|Q)\right)}$, the following alternative bound holds:
\begin{align}
    V\geq\frac{1}{2}\gamma L\left(\mathbb{E}_{\theta_{0}}[\hat{\phi}],\mathbb{E}_{\theta_{1}}[\hat{\phi}]\right)\left(1-\sqrt{1-e^{-\mathrm{KL}(\mathcal{P}_{\theta_{0}}\|\mathcal{P}_{\theta_{1}})}}\right).
    \label{eq:Mprime_BretagnolleHuber}
\end{align}
Consider a steady-state Markov process and suppose that the parameters $\theta_0$ and $\theta_1$ correspond to forward and backward dynamics, respectively. 
The Kullback-Leibler divergence between forward and backward path probabilities is the entropy production:
\begin{align}
    \Sigma=\mathrm{KL}(\mathcal{P}[\Gamma_{\tau}]\|\mathcal{P}[\Gamma^{\dagger}_{\tau}]),
    \label{eq:entropy_production_KL}
\end{align}
where $\Gamma_\tau^\dagger$ is the time reversal of $\Gamma_\tau$. 
We consider the following observable, which is odd under time reversal:
\begin{align}
    \hat{\phi}_{\rightleftarrows}(\Gamma_{\tau})=-\hat{\phi}_{\rightleftarrows}(\Gamma^{\dagger}_{\tau}).
    \label{eq:phi_time_reversal}
\end{align}
Then, we obtain the following relation:
\begin{align}
\frac{\mathbb{E}\left[\left|\hat{\phi}_{\rightleftarrows}(\Gamma_{\tau})-\mathbb{E}[\hat{\phi}_{\rightleftarrows}]\right|^{p}\right]}{\left|\mathbb{E}[\hat{\phi}_{\rightleftarrows}]\right|^{p}}\geq1-\sqrt{1-e^{-\Sigma}},\label{eq:generalized_point_TUR}
\end{align}
which holds under steady-state conditions. 
The derivation is shown in Appendix~\ref{sec:derivation_point_TUR}. 
The thermodynamic uncertainty relation for the time-reversal observable in Eq.~\eqref{eq:phi_time_reversal} was shown in Refs.~\cite{Hasegawa:2019:FTUR,Timpanaro:2019:EFTTUR}.
Although the derived bound holds for a general $b$-metric loss function, it is weaker than those in Refs.~\cite{Hasegawa:2019:FTUR,Timpanaro:2019:EFTTUR}.

\section{Conclusion\label{sec:conclusion}}
In this study, we considered statistical decision theory in stochastic thermodynamics.
We focused on the minimax risk, the smallest achievable expected loss under the worst-case parameter choice, even with an optimal estimator.
We derived lower bounds on the minimax risk using Le Cam's two-point method. 
As an example of the derived bounds,
we studied classification and distribution estimation problems in a Markov process and showed that the minimax risk provides lower bounds given by the dynamical activity.
Moreover, by considering particular dynamics, we demonstrated that the resulting bounds reduce to uncertainty relations comprising dynamical activity or entropy production.

\begin{acknowledgments}

This work was supported by JSPS KAKENHI Grant Numbers JP26K02998, JP24K03008, and JP26K06158.

ChatGPT 5.6 was used to assist with checking for errors.
All scientific ideas, analyses, calculations, and conclusions are those of the author.

\end{acknowledgments}

\appendix

\section{Le Cam's two-point method \label{sec:lecam_derivation}}

Here, we review the derivation of Le Cam's two-point method \cite{LeCam:1973:dimensionality,Raginsky:2013:SLT,Wainwright:2019:HighDimensionalStatistics}. We follow the derivation presented in Ref.~\cite{Huang:2025:LeCam}.
The quantity $\max_{\theta\in\Theta}\mathbb{E}_{\theta}[L(\theta,\hat{\theta})]$ has the following lower bound:
\begin{align}
&\max_{\theta\in\Theta}\mathbb{E}_{\theta}[L(\theta,\hat{\theta})]\nonumber\\&\ge\max_{\theta\in\{\theta_{0},\theta_{1}\}}\mathbb{E}_{\theta}[L(\theta,\hat{\theta})]\nonumber\\&\ge\sum_{X}\frac{L\left(\theta_{0},\hat{\theta}(x)\right)P_{\theta_{0}}(X)+L\left(\theta_{1},\hat{\theta}(X)\right)P_{\theta_{1}}(X)}{2}\nonumber\\&\ge\sum_{X}\frac{L\left(\theta_{0},\hat{\theta}(X)\right)+L\left(\theta_{1},\hat{\theta}(X)\right)}{2}\min\left(P_{\theta_{0}}(X),P_{\theta_{1}}(X)\right)\nonumber\\&\ge\frac{\gamma L(\theta_{0},\theta_{1})}{2}\sum_{x}\min\left(P_{\theta_{0}}(X),P_{\theta_{1}}(X)\right)\nonumber\\&=\frac{\gamma L(\theta_{0},\theta_{1})}{2}\left(1-\mathrm{TV}\left(P_{\theta_{0}},P_{\theta_{1}}\right)\right).
    \label{eq:LeCam_derivation}
\end{align}
From the fourth to the fifth line, Eq.~\eqref{eq:pseudo_metric_condition1} is used. 
From the fifth to the sixth line, the following relation was used:
\begin{align}
    \min (a, b)=\frac{a+b}{2}-\frac{1}{2}|b-a|.
    \label{eq:min_as_abs}
\end{align}
The final line of Eq.~\eqref{eq:LeCam_derivation} is Le Cam's two-point method.

\section{Derivation of Eq.~\eqref{eq:M_lowerbound_TV_thermo2}\label{sec:derivation_M_lowerbound_TV_thermo2}}

We show the derivation of Eq.~\eqref{eq:M_lowerbound_TV_thermo2}. 
The derivation is almost the same as in Section~\ref{sec:lecam_derivation}.
The quantity $\max_{\theta\in\Theta}\mathbb{E}_{\theta}[L(\mathbb{E}_{\theta}[\hat{\phi}],\hat{\phi}(\Gamma_{\tau}))]$ has the following lower bound:
\begin{align}
&\max_{\theta\in\Theta}\mathbb{E}_{\theta}[L(\mathbb{E}_{\theta}[\hat{\phi}],\hat{\phi}(\Gamma_{\tau}))]\nonumber\\&\ge\max_{\theta\in\{\theta_{0},\theta_{1}\}}\mathbb{E}_{\theta}[L(\mathbb{E}_{\theta}[\hat{\phi}],\hat{\phi}(\Gamma_{\tau}))]\nonumber\\&\ge\sum_{\Gamma_{\tau}}\frac{L(\mathbb{E}_{\theta_{0}}[\hat{\phi}],\hat{\phi}(\Gamma_{\tau}))\mathcal{P}_{\theta_{0}}[\Gamma_{\tau}]+L(\mathbb{E}_{\theta_{1}}[\hat{\phi}],\hat{\phi}(\Gamma_{\tau}))\mathcal{P}_{\theta_{1}}[\Gamma_{\tau}]}{2}\nonumber\\&\ge\sum_{\Gamma_{\tau}}\frac{L(\mathbb{E}_{\theta_{0}}[\hat{\phi}],\hat{\phi}(\Gamma_{\tau}))+L(\mathbb{E}_{\theta_{1}}[\hat{\phi}],\hat{\phi}(\Gamma_{\tau}))}{2}\nonumber\\&\times\min\left(\mathcal{P}_{\theta_{0}}[\Gamma_{\tau}],\mathcal{P}_{\theta_{1}}[\Gamma_{\tau}]\right)\nonumber\\&\ge\frac{\gamma L(\mathbb{E}_{\theta_{0}}[\hat{\phi}],\mathbb{E}_{\theta_{1}}[\hat{\phi}])}{2}\sum_{\Gamma_{\tau}}\min\left(\mathcal{P}_{\theta_{0}}[\Gamma_{\tau}],\mathcal{P}_{\theta_{1}}[\Gamma_{\tau}]\right)\nonumber\\&=\frac{\gamma L(\mathbb{E}_{\theta_{0}}[\hat{\phi}],\mathbb{E}_{\theta_{1}}[\hat{\phi}])}{2}\left(1-\mathrm{TV}\left(\mathcal{P}_{\theta_{0}},\mathcal{P}_{\theta_{1}}\right)\right).
\label{eq:LeCam_phi_derivation}
\end{align}
The final line in Eq.~\eqref{eq:LeCam_phi_derivation} is Eq.~\eqref{eq:M_lowerbound_TV_thermo2} in the main text.

\section{Derivation of Eq.~\eqref{eq:generalized_KUR}\label{sec:derivation_gKUR}}

Using the Pinsker inequality \cite{Pinsker:1964:Book},
from Eq.~\eqref{eq:M_lowerbound_TV_thermo2}, we have
\begin{align}
    V\geq\frac{1}{2}\gamma L\left(\mathbb{E}_{\theta_{0}}[\hat{\phi}],\mathbb{E}_{\theta_{1}}[\hat{\phi}]\right)\left(1-\sqrt{\frac{1}{2}\mathrm{KL}(\mathcal{P}_{\theta_{0}}\|\mathcal{P}_{\theta_{1}})}\right),
    \label{eq:Mprime_KL_appendix}
\end{align}
It is known that the Kullback-Leibler divergence and the Fisher information $\mathcal{I}(\theta)$ are related via
\begin{align}
    \mathrm{KL}(\mathcal{P}_{\theta}\|\mathcal{P}_{\theta+\Delta\theta})\simeq\frac{1}{2}\Delta\theta^{2}\mathcal{I}(\theta),
    \label{eq:KL_Fisher}
\end{align}
where $0<\Delta \theta \ll 1$. 
Moreover, when the transition rate is parametrized as Eq.~\eqref{eq:W_parametrized_theta}, it is known that the Fisher information becomes the dynamical activity:
\begin{align}
    \mathcal{I}(\theta)=\frac{\tau}{\theta}\sum_{\mu,\nu\,(\mu\ne\nu)}W_{\nu\mu}P_{\mathrm{ss}}(\mu).
    \label{eq:Fisher_to_DA}
\end{align}
Moreover, when the loss function is $|\theta-\theta^\prime|^p$, we have
\begin{align}
L(\mathbb{E}_{\theta}[\hat{\phi}_{\sharp}],\mathbb{E}_{\theta+\Delta\theta}[\hat{\phi}_{\sharp}])&=\left|\mathbb{E}_{\theta+\Delta\theta}[\hat{\phi}_{\sharp}]-\mathbb{E}_{\theta}[\hat{\phi}_{\sharp}]\right|^{p}\nonumber\\&=\left|\left(1+\frac{\Delta\theta}{\theta}\right)\mathbb{E}_{\theta}[\hat{\phi}_{\sharp}]-\mathbb{E}_{\theta}[\hat{\phi}_{\sharp}]\right|^{p}\nonumber\\&=\left(\frac{\Delta\theta}{\theta}\right)^{p}\left|\mathbb{E}_{\theta}[\hat{\phi}_{\sharp}]\right|^{p}.
    \label{eq:E_difference}
\end{align}
Therefore, for $\theta=1$, the right-hand side of Eq.~\eqref{eq:Mprime_KL_appendix} becomes
\begin{align}
V\geq\frac{\gamma}{2}\left(\Delta\theta\right)^{p}\left|\mathbb{E}_{\theta}[\hat{\phi}_{\sharp}]\right|^{p}\left(1-\frac{1}{2}\Delta\theta\sqrt{\mathfrak{a}_{\mathrm{ss}}\tau}\right),
    \label{eq:Mprime_steadystate_Deltatheta}
\end{align}
We can freely choose $\Delta \theta$ as long as $0<\Delta \theta \ll 1$. Therefore, we maximize the right-hand side of Eq.~\eqref{eq:Mprime_steadystate_Deltatheta} as a function of $\Delta \theta$. 
When $p>1$, the maximum is achieved at $\Delta\theta=2p/(\sqrt{\mathfrak{a}_{\mathrm{ss}}\tau}(p+1))$. 
This $\Delta \theta$ satisfies $\Delta \theta \ll 1$ for $\tau \to \infty$, so it is acceptable. 
\begin{align}
    V\ge\frac{\gamma}{2}\left|\mathbb{E}_{\theta}[\hat{\phi}_{\sharp}]\right|^{p}\frac{2^{p}p^{p}}{(\mathfrak{a}_{\mathrm{ss}}\tau)^{p/2}(1+p)^{p+1}}.
    \label{eq:Mprime_steadystate_counting}
\end{align}
Since $\Delta \theta \to 0$ for $\tau \to \infty$, 
$\mathbb{E}_{\theta+\Delta\theta}[|\hat{\phi}_{\sharp}(\Gamma_{\tau})-\mathbb{E}_{\theta+\Delta\theta}[\hat{\phi}_{\sharp}]|^{p}]=\mathbb{E}_{\theta}[|\hat{\phi}_{\sharp}(\Gamma_{\tau})-\mathbb{E}_{\theta}[\hat{\phi}_{\sharp}]|^{p}]$ for $p > 1$. 
With $\gamma = 2^{1-p}$,
using Eq.~\eqref{eq:Mprime_steadystate_counting},
we obtain
\begin{align}
    \frac{\mathbb{E}[|\hat{\phi}_{\sharp}(\Gamma_{\tau})-\mathbb{E}[\hat{\phi}_{\sharp}]|^{p}]}{\left|\mathbb{E}[\hat{\phi}_{\sharp}]\right|^{p}}\ge\frac{p^{p}}{(\mathfrak{a}_{\mathrm{ss}}\tau)^{p/2}(1+p)^{p+1}},
    \label{eq:generalized_KUR_appendix}
\end{align}
which is Eq.~\eqref{eq:generalized_KUR} in the main text. 
Here, we used $\mathbb{E}_{\theta=1}[\hat{\phi}_\sharp] = \mathbb{E}[\hat{\phi}_\sharp]$.

\section{Derivation of Eq.~\eqref{eq:distribution_bound}\label{sec:distribution_derivation_appendix}}

Here, we show the derivation of Eq.~\eqref{eq:distribution_bound}. 
From Eq.~\eqref{eq:thermo_M_KL_lowerbound}, the following relation holds:
\begin{align}
&\min_{\hat{P}_{\mathrm{ss}}}\max_{\mathcal{P}\in\mathscr{P}_{\mathrm{ss}}}\mathbb{E}_\mathcal{P}\left[L(\hat{P}_{\mathrm{ss}},P_{\mathrm{ss}}(\mathcal{P}))\right]\nonumber\\&\ge\frac{1}{2\sqrt{2}}\frac{\theta}{2+\theta}\left(1-\sqrt{\frac{1}{2}\mathrm{KL}(\mathcal{P}_{\theta_{0}}\|\mathcal{P}_{\theta_{1}})}\right)\nonumber\\&\simeq\frac{1}{2\sqrt{2}}\frac{\theta}{2+\theta}\left(1-\frac{\theta}{2}\sqrt{\mathcal{I}(\theta=0)}\right)\,\,\,(\theta\ll1).
\label{eq:minimax_distribution_lowerbound_app}
\end{align}
In the last line, we followed the same argument as in Section~\ref{sec:derivation_gKUR}. 
Since we are assuming the $\tau \to \infty$ limit, the initial distribution does not affect the Fisher information $\mathcal{I}(\theta)$. 
Similarly to Eq.~\eqref{eq:Fisher_to_DA},
the Fisher information $\mathcal{I}(\theta)$ reduces to the dynamical activity.
However, since the parameter $\theta$ enters only in the transition $2\to1$, for $\tau \to \infty$, we have
\begin{align}
    \mathcal{I}(\theta = 0)=\tau P_{\mathrm{ss}}(2)W_{12}=\frac{\tau}{2},
    \label{eq:Itheta_12}
\end{align}
which is the activity corresponding to the transition $2\to 1$. 
Next, we maximize Eq.~\eqref{eq:minimax_distribution_lowerbound_app} with respect to $\theta$. 
We find that the maximum is achieved at
\begin{align}
    \theta=2\left(-1+\frac{\sqrt{\sqrt{2\tau}+\tau}}{\sqrt{\tau}}\right),
    \label{eq:theta_max_at}
\end{align}
where the lower bound becomes
\begin{align}
    &\min_{\hat{P}_{\mathrm{ss}}}\max_{\mathcal{P}\in\mathscr{P}_{\mathrm{ss}}}\mathbb{E}_{\mathcal{P}}\left[L(\hat{P}_{\mathrm{ss}},P_{\mathrm{ss}}(\mathcal{P}))\right]\nonumber\\&\ge\frac{1}{2\sqrt{2}}\left(1+\sqrt{2\tau}-\sqrt{2}\sqrt{\sqrt{2\tau}+\tau}\right).
    \label{eq:distribution_bound_app}
\end{align}
Recall that the following relation holds for $\tau \to \infty$:
\begin{align}
    1+\sqrt{2\tau}-\sqrt{2}\sqrt{\sqrt{2\tau}+\tau}\to\frac{1}{2\sqrt{2\tau}}.
    \label{eq:tau_inf_limit_lowerbound}
\end{align}
From Eq.~\eqref{eq:tau_inf_limit_lowerbound}, we obtain Eq.~\eqref{eq:distribution_bound} in the main text. 

\section{Derivation of Eq.~\eqref{eq:generalized_point_KUR}\label{sec:derivation_pointKUR}}

We consider two parameters $\theta_0$ and $\theta_1$. 
The transition rate $W_{\nu\mu}$ ($\mu \ne \nu$) is parametrized as
\begin{align}
    W_{\nu\mu}(\theta)=\begin{cases}
W_{\nu\mu} & \theta=\theta_{1}\\
0 & \theta=\theta_{0}
\end{cases}.
\label{eq:Wnumu_theta0_theta1}
\end{align}
From Eq.~\eqref{eq:M_lowerbound_TV_thermo2}, we have
\begin{align}
    V\geq\frac{1}{2}\gamma L\left(\mathbb{E}_{\theta_{0}}[\hat{\phi}_{\circ}],\mathbb{E}_{\theta_{1}}[\hat{\phi}_{\circ}]\right)\left(1-\sqrt{1-e^{-\tau(\mathfrak{a}_{\theta_{0}}(0)+\mathfrak{a}_{\theta_{1}}(0))}}\right).
    \label{eq:Mprime_phi0_bound_appendix}
\end{align}
From the condition of Eq.~\eqref{eq:nojump_trajectory} and the fact that the dynamics corresponding to $\theta_0$ is empty, we have
\begin{align}
    \mathbb{E}_{\theta_{0}}[\hat{\phi}_{\circ}]&=0,\label{eq:E_phi0_zero}\\\mathfrak{a}_{\theta_{0}}(0)&=0.
    \label{eq:a_theta0_zero}
\end{align}
Equation~\eqref{eq:Mprime_phi0_bound_appendix} becomes
\begin{align}
    V\geq\frac{1}{2}\gamma L\left(0,\mathbb{E}_{\theta_{1}}[\hat{\phi}_{\circ}]\right)\left(1-\sqrt{1-e^{-\tau\mathfrak{a}_{\theta_{1}}(0)}}\right).
    \label{eq:Mprime_phi0_bound_appendix2}
\end{align}
The left-hand side of Eq.~\eqref{eq:Mprime_phi0_bound_appendix2} is
\begin{align}
&\max\left[\mathbb{E}_{\theta_{0}}[L(\mathbb{E}_{\theta_{0}}[\hat{\phi}_{\circ}],\hat{\phi}(\Gamma_{\tau}))],\mathbb{E}_{\theta_{1}}[L(\mathbb{E}_{\theta_{1}}[\hat{\phi}_{\circ}],\hat{\phi}_{\circ}(\Gamma_{\tau}))]\right]\nonumber\\&=\max\left[\mathbb{E}_{\theta_{0}}[L(0,0)],\mathbb{E}_{\theta_{1}}[L(\mathbb{E}_{\theta_{1}}[\hat{\phi}_{\circ}],\hat{\phi}_{\circ}(\Gamma_{\tau}))]\right]\nonumber\\&=\mathbb{E}_{\theta_{1}}[L(\mathbb{E}_{\theta_{1}}[\hat{\phi}_{\circ}],\hat{\phi}_{\circ}(\Gamma_{\tau}))]\nonumber\\&=\mathbb{E}[L(\mathbb{E}[\hat{\phi}_{\circ}],\hat{\phi}_{\circ}(\Gamma_{\tau}))].
    \label{eq:max_theta0_theta1}
\end{align}
Combining Eqs.~\eqref{eq:Mprime_phi0_bound_appendix2} and \eqref{eq:max_theta0_theta1}, we obtain Eq.~\eqref{eq:generalized_point_KUR} in the main text. 

\section{Derivation of Eq.~\eqref{eq:generalized_point_TUR}
\label{sec:derivation_point_TUR}
}

We consider two parameters $\theta_0$ and $\theta_1$. 
The path probability is parametrized as
\begin{align}
    \mathcal{P}_{\theta}[\Gamma_{\tau}]=\begin{cases}
\mathcal{P}[\Gamma_{\tau}] & \theta=\theta_{0}\\
\mathcal{P}[\Gamma^{\dagger}_{\tau}] & \theta=\theta_{1}
\end{cases}.
\label{eq:mathcalP_parametrize}
\end{align}
It is known that the Kullback-Leibler divergence between forward and backward probabilities is the entropy production:
\begin{align}
    \Sigma=\mathrm{KL}\left(\mathcal{P}[\Gamma_{\tau}]\|\mathcal{P}[\Gamma^{\dagger}_{\tau}]\right).
    \label{eq:KL_entropy_production_appendix}
\end{align}
Moreover, the time-reversal observable satisfies
\begin{align}
\mathbb{E}_{\theta_{1}}[\hat{\phi}_{\rightleftarrows}]&=\sum_{\Gamma_{\tau}}\mathcal{P}[\Gamma^{\dagger}_{\tau}]\hat{\phi}_{\rightleftarrows}(\Gamma_{\tau})=\sum_{\Gamma_{\tau}}\mathcal{P}[\Gamma_{\tau}]\hat{\phi}_{\rightleftarrows}(\Gamma^{\dagger}_{\tau})\nonumber\\&=-\sum_{\Gamma_{\tau}}\mathcal{P}[\Gamma_{\tau}]\hat{\phi}_{\rightleftarrows}(\Gamma_{\tau})=-\mathbb{E}_{\theta_{0}}[\hat{\phi}_{\rightleftarrows}]\nonumber\\&=-\mathbb{E}[\hat{\phi}_{\rightleftarrows}],
    \label{eq:time_reversal_observable}
\end{align}
where $\mathbb{E}[\cdot] = \mathbb{E}_{\theta_0}[\cdot]$. 
Substituting Eqs.~\eqref{eq:KL_entropy_production_appendix} and \eqref{eq:time_reversal_observable} into Eq.~\eqref{eq:Mprime_BretagnolleHuber}, we obtain
\begin{align}
    V\geq\frac{1}{2}\gamma L\left(\mathbb{E}[\hat{\phi}_{\rightleftarrows}],-\mathbb{E}[\hat{\phi}_{\rightleftarrows}]\right)\left(1-\sqrt{1-e^{-\Sigma}}\right).
    \label{eq:Mprime_current_appendix}
\end{align}
Next, we evaluate the left-hand side of Eq.~\eqref{eq:Mprime_current_appendix}.
Recall that
\begin{align}
    &\mathbb{E}_{\theta_{1}}[L(\mathbb{E}_{\theta_{1}}[\hat{\phi}_{\rightleftarrows}],\hat{\phi}_{\rightleftarrows}(\Gamma_{\tau}))]\nonumber\\&=\sum_{\Gamma_{\tau}}\mathcal{P}[\Gamma^{\dagger}_{\tau}]L(\mathbb{E}_{\theta_{1}}[\hat{\phi}_{\rightleftarrows}],\hat{\phi}_{\rightleftarrows}(\Gamma_{\tau}))\nonumber\\&=\sum_{\Gamma_{\tau}}\mathcal{P}[\Gamma_{\tau}]L(-\mathbb{E}[\hat{\phi}_{\rightleftarrows}],-\hat{\phi}_{\rightleftarrows}(\Gamma_{\tau}))\nonumber\\&=\mathbb{E}\left[L(-\mathbb{E}[\hat{\phi}_{\rightleftarrows}],-\hat{\phi}_{\rightleftarrows}(\Gamma_{\tau}))\right].
    \label{eq:expectation_Loss}
\end{align}
From Eq.~\eqref{eq:expectation_Loss}, we obtain
\begin{align}
&\max\left[\mathbb{E}_{\theta_{0}}[L(\mathbb{E}_{\theta_{0}}[\hat{\phi}_{\rightleftarrows}],\hat{\phi}_{\rightleftarrows}(\Gamma_{\tau}))],\mathbb{E}_{\theta_{1}}[L(\mathbb{E}_{\theta_{1}}[\hat{\phi}_{\rightleftarrows}],\hat{\phi}_{\rightleftarrows}(\Gamma_{\tau}))]\right]\nonumber\\&=\max\left[\mathbb{E}[L(\mathbb{E}[\hat{\phi}_{\rightleftarrows}],\hat{\phi}_{\rightleftarrows}(\Gamma_{\tau}))],\mathbb{E}\left[L(-\mathbb{E}[\hat{\phi}_{\rightleftarrows}],-\hat{\phi}_{\rightleftarrows}(\Gamma_{\tau}))\right]\right].
    \label{eq:max_theta0_theta1_current}
\end{align}
Specifically, we consider the loss function given in Eq.~\eqref{eq:Lp_def}. 
Since $L(\mathbb{E}_{\theta_{1}}[\hat{\phi}_{\rightleftarrows}],\hat{\phi}_{\rightleftarrows}(\Gamma_{\tau}))=|-\mathbb{E}[\hat{\phi}_{\rightleftarrows}]+\hat{\phi}_{\rightleftarrows}(\Gamma_{\tau})|^{p}$, we obtain
\begin{align}
    \frac{\mathbb{E}\left[\left|\mathbb{E}[\hat{\phi}_{\rightleftarrows}]-\hat{\phi}_{\rightleftarrows}(\Gamma_{\tau})\right|^{p}\right]}{|\mathbb{E}[\hat{\phi}_{\rightleftarrows}]|^{p}}\geq1-\sqrt{1-e^{-\Sigma}},
    \label{eq:generalized_point_TUR_appendix}
\end{align}
which is Eq.~\eqref{eq:generalized_point_TUR} in the main text. 

\section{Le Cam lower bound in Gaussian mean\label{sec:normal_mean}}

As a well-known pedagogical example, let us consider parameter estimation for the mean of a Gaussian distribution when the variance is known \cite{Polyanskiy:2025:InformationTheory}. 
Let $\xi$ be the mean of the Gaussian distribution and $\hat{\xi}$ be its estimator. 
From Eq.~\eqref{eq:thermo_M_KL_lowerbound}, the following relation holds:
\begin{align}
    \max_{\xi}\mathbb{E}_{\xi}[(\hat{\xi}-\xi)^{2}]\ge\frac{(\xi_{1}-\xi_{2})^{2}}{4}\left(1-\sqrt{\frac{1}{2}\mathrm{KL}\left(P_{\xi_{1}}\|P_{\xi_{2}}\right)}\right),
    \label{eq:Emu_lowerbound}
\end{align}
where $\xi_1$ and $\xi_2$ are arbitrary mean values. 
Suppose that we choose $\xi_2 - \xi_1=\delta$. 
Given $n$ independent data points, the Kullback-Leibler divergence for the Gaussian case is given by
\begin{align}
    \mathrm{KL}(P_{\xi_{1}}\|P_{\xi_{2}})=\frac{n\delta^{2}}{2\sigma^{2}}.
    \label{eq:Gaussian_KL}
\end{align}
Therefore, the minimax risk has the following lower bound:
\begin{align}
    \max_{\xi}\mathbb{E}_{\xi}[(\hat{\xi}-\xi)^{2}]&\ge\frac{\delta^{2}}{4}\left(1-\frac{\delta\sqrt{n}}{2\sigma}\right).
    \label{eq:Emu_lowerbound2}
\end{align}
Since we can choose $\delta$ so as to maximize the right-hand side of Eq.~\eqref{eq:Emu_lowerbound2}, the maximum is achieved at $\delta = 4\sigma/(3\sqrt{n})$:
\begin{align}
    \max_{\xi}\mathbb{E}_{\xi}[(\hat{\xi}-\xi)^{2}]\ge\frac{4\sigma^{2}}{27n}=O\left(\frac{\sigma^{2}}{n}\right).
    \label{eq:Gaussian_minimax}
\end{align}
For comparison, we consider the Cram\'er-Rao inequality in the same setting. The Fisher information of the Gaussian mean is known to be
\begin{align}
    \mathcal{I}(\xi)=\frac{n}{\sigma^{2}}.
    \label{eq:Imu_Gaussian}
\end{align}
Therefore, the variance of an unbiased estimator has the following lower bound:
\begin{align}
    \mathrm{Var}[\hat{\xi}]\ge\frac{1}{\mathcal{I}(\xi)}=\frac{\sigma^{2}}{n}.
    \label{eq:Gaussian_mean_CramerRao}
\end{align}
In Eq.~\eqref{eq:Gaussian_mean_CramerRao}, the lower bound is given by $\sigma^2/n$, while that of Le Cam's method is smaller by a constant factor. 
This shows that Le Cam's method is weaker than the Cram\'er-Rao inequality.
However, it is important to note that the estimator considered in Le Cam's method is not restricted to unbiased estimators.
This contrasts with the Cram\'er-Rao inequality, in which estimators must be unbiased.

\end{document}